\documentclass[12pt]{iopart}

\usepackage{graphicx}
\usepackage{harvard}
\newcommand\bav{\begin{array}{c}}  	
\newcommand\bam{\begin{array}{cc}}  	
\newcommand\ea{\end{array}}  
\newcommand\beq{\begin{equation}} 
\newcommand\eeq{\end{equation}}  
\newcommand\bea{\begin{eqnarray}}  
\newcommand\eea{\end{eqnarray}}  
\newcommand\beas{\begin{eqnarray*}}  
\newcommand\eeas{\end{eqnarray*}}  
\newcommand\Rey{\mbox{Re}}

\begin{document}

\title[Small scale exact coherent structures]{Small scale exact coherent structures at large Reynolds numbers in plane Couette flow}  
  
\author{Bruno Eckhardt}
\address{Fachbereich Physik, Philipps-Universt\"at 
Marburg, Renthof 6, \\ 35032 Marburg, Germany} 
\ead{bruno.eckhardt@physik.uni-marburg.de}

\author{Stefan Zammert}
\address{Laboratory for Aero and Hydrodynamics, TU Delft, 2682 Delft, 
The Netherlands}

\vspace{10pt}
\begin{indented}
\item[]July 2017
\end{indented}

\begin{abstract}
The transition to turbulence in plane Couette flow and several other shear flows
is connected with saddle node bifurcations in which fully 3-d, nonlinear solutions,
so-called exact coherent states (ECS),
to the Navier-Stokes equation appear. As the Reynolds number
increases, the states undergo secondary bifurcations and their time-evolution
becomes increasingly more complex. Their spatial complexity, in contrast, remains
limited so that these states cannot contribute to the spatial complexity and 
cascade to smaller scales expected for higher Reynolds numbers. We here present families
of scaling ECS that exist on ever smaller scales as the Reynolds number 
is increased. We focus in particular on two such families for plane
Couette flow, one centered near the midplane and the other close to a wall. 
We discuss their scaling and localization properties and the bifurcation diagrams.
All solutions are localized in the wall-normal direction. In the spanwise and 
downstream direction, they are either periodic or localized as well.  The family of 
scaling ECS localized near a wall is reminiscent of attached eddies, 
and indicates how self-similar ECS can contribute to the formation of boundary layer profiles.
\end{abstract}


\maketitle

\section{Introduction}
Pipe flow and various other parallel and non-parallel shear flows show
a transition to turbulence that is not connected to a linear instability of
the laminar profile \cite{Grossmann:2000}.
The transition can be triggered by finite amplitude bifurcations
and the new states that emerge are spatially and temporally fluctuating.
The origin of the transition and the subsequent dynamics
cannot be understood within linear approximations but require that 
the nonlinearity of the Navier-Stokes equation is taken into account.
Underlying the complex spatio-temporal patterns are exact coherent
states (ECS) \cite{Waleffe:1998wk,Waleffe:2001wu}, 
i.e. velocity fields that are solutions to the Navier-Stokes equation with
a relatively simple temporal dynamics: they can be fixed points of the equations of 
motion, travelling waves or more complex relative periodic orbits
\cite{Eckhardt:2007ix,Kerswell:2005ir,Eckhardt:2007ka}.
ECS provide nuclei for the formation of turbulence.
They typically appear in saddle-node bifurcations \cite{Mellibovsky:2011gq}
and then undergo sequences of secondary bifurcations that give temporally
complex dynamics \cite{Kreilos:2012bd,Avila:2013jq,Zammert:2015jg}.
Crisis bifurcations can change the dynamics from persistent
to transient, and collisions between different coherent structures can set up
a network that sustains long-lived turbulent dynamics
\cite{Hof:2006ab,Hof:2008cb,Avila:2010ep,Schneider:2010gv,Kreilos:2014ew}.


The bifurcations just described follow the patterns familiar from the various 
routes to chaos and can explain the temporally complex dynamics
\cite{Eckmann:1981kw,Ott:2002wz}. In order to realize the distribution 
of energy to ever smaller scales that are the  
%
hallmark of fully developed turbulence \cite{Frisch:1995wl} mechanisms
that create structures on smaller scales are required.
%
Steps towards developed turbulence are described in the studies of \cite{Kawahara:2001ft}
where it is shown that ECS can capture some of the turbulent dynamics,
and \cite{vanVeen:2009fm}, 
where coherent structures for models of homogeneous turbulence are described. 
In all cases the Reynolds numbers are moderate and the structures remain large-scale
in the sense that they extend all the way across the available volume. 
The examples presented below belong to families of 
states that can be scaled to ever finer spatial scales as the Reynolds number 
increases. 

All ECS are fully three-dimensional: all velocity components are active and
they vary in all three directions. Simpler structures, e.g. with translational
invariance in the downstream direction, decay
\cite{Moffatt:1990fb}.
Many of them share relatively stable relations between their height, width, 
and downstream periodicity:
if $H$ denotes the height, then the width of the structures is about $\pi H$ and the
downstream wavelength is about $2\pi H$. For plane Couette flow, the exact optimal
relations are documented in 
\cite{Clever:1997tq,Waleffe:2003hh}, 
and the estimates for pipe flow are
given in 
\cite{Faisst:2003hd,Eckhardt:2008jv,Pringle:2009fe}.
Similar results are available for plane Poiseuille flow, 
though the optimal wavelength described in 
\cite{Zammert:2016fk}
shows that there is some variability in the optimal ratios.  All ECS just described
span across the entire height of the shear flow. 

An approach to finding smaller structures is suggested by the behaviour of 
states in  pipe flow 
\cite{Faisst:2003hd,Eckhardt:2008jv,Pringle:2009fe}: 
as number of vortices along the circumference
increases, they move closer to the walls and also their downstream wavelength
decreases. Apparently, the vortices try to maintain the geometric relations as they
become narrower.
This observation suggests that smaller structures can be obtained
by scaling structures in all three directions, and specifically by prescribing
the spanwise wavelength, so that the extension of the states in the normal
and downstream direction has to adjust to the prescribed widths. 
\cite{Hall:2010cn} noted such a scaling for states in their asymptotic 
expansion for high Reynolds numbers and
\cite{Deguchi:2015gu}
showed that this is one of the scalings inherent in this expansion.
Here, we employ this scaling to find approximate rescaled states that are then
refined using a Newton step to arrive at a state that is an ECS of the full
Navier-Stokes equation at a prescribed Reynolds number.
We apply this to several states from plane Couette flow, trace them  to 
high Reynolds numbers, and show their bifurcation and scaling properties. 
Of particular interest are
a set of structures that are localized near the walls and which can be viewed
as ECS that may support the popular image of
boundary layers being carried by a hierarchy of eddies attached
to the walls \cite{Townsend:1980uj,Perry:1991ab,Perry:1994ab}.

In the next section, we present the scaling ansatz for plane Couette flow.
In section 3, we discuss the properties of states. We first focus
on states that are periodic in the spanwise and downstream direction, and
that are localized in the center (section 3.1) and near a wall (section 3.2).
We then describe their bifurcation and scaling structure (section 3.3), 
their localization in the normal direction (section 3.4) and their
stability properties (section 3.5).
States that are also localized in the downstream or spanwise direction
are described in section 4. 
Conclusions are given in section 5.

\section{Scaling in plane Couette flow}
The flow we consider here is plane Couette flow, the flow between
two parallel plates moving relative to each other. With $x$ the downstream
direction, $y$ the normal direction, and $z$ the spanwise direction, the 
laminar profile is $\mathbf{u}_0(x,y,z) = S y$ with the shear $S=U_0/d$ for plates at
$y=\pm d$ that move with velocity $\pm U_0$. 
Deviations $\mathbf{u}$ from the laminar profile then satisfy the equation
\begin{equation}
\partial_t \mathbf{u} + (Sy \mathbf{e}_{x} \cdot\nabla)\mathbf{u} + 
(\mathbf{u} \cdot\nabla) Sy\mathbf{e}_{x}  +  ( \mathbf{u} \cdot\nabla) \mathbf{u} 
+ \nabla p
=  \nu \Delta  \mathbf{u} \,.
\end{equation}
with $\nu$ the kinematic viscosity.
For stationary states, $\partial_t \mathbf{u}=0$ and only the spatial degrees
of freedom remain.
For states moving with a constant phase velocity $c$ in the downstream
direction, transition to a comoving frame $\tilde x = x - c t$ gives
$\partial_t \mathbf{u} = - c \partial_x \mathbf{u}$ and a time-independent
equation in $\tilde x$.

Let $\mathbf{u}_0(\mathbf{x})$ be a solution to the stationary
equation for a viscosity $\nu_0$. Then the scaled velocity field 
%
%
\beq
\mathbf{u}_{\lambda}(\mathbf{x})= \mathbf{u}_0(\lambda \mathbf{x}) /\lambda
\label{u_scaled}
\eeq
satisfies
\begin{equation}
-c^* \partial_x \mathbf{u}_0 + 
(Sy^* \mathbf{e}_{x} \cdot\nabla)\mathbf{u}_0+ 
(\mathbf{u}_{0}\cdot\mathbf{e}_y)S \mathbf{e}_{x}  +  
(\mathbf{u}_{0} \cdot\nabla) \mathbf{u}_{0} + \nabla p^*
=  \nu^{*} \Delta  \mathbf{u}_0  
\label{eq:fixedpoint}
\end{equation}
in the scaled coordinates $\mathbf{x}^* = \lambda \mathbf{x}$ with
$c^* = c\lambda $ and $\nu^*= \nu/\lambda^2$. That is to say,
$\mathbf{u}_\lambda$ is a solution at the modified viscosity $\nu^*$.
With this transformation, we also have to  adjust the walls, and they move outwards 
at the same rate, $d^* = d\lambda$. However, if the state is localized in the
normal direction, the velocity fields will decay towards the walls and the 
specific location of the walls will only have a small influence on the state.
By the above heuristics,
the state will be localized in the normal direction if the spanwise and/or downstream periodicity 
are small compared to the initial distance between the walls.

To see the scaling in Reynolds number, we define $\Rey=(Sd)d/\nu$, so that
the rescaled state $\mathbf{u}_{\lambda}$  are equilibrium states for 
\beq
\Rey^{*}=\lambda^{2} \Rey_{0}.
\eeq
or, alternatively, that a solution at Reynolds number $\Rey^{*}$ can be obtained 
with the rescaling
\beq
\lambda=\sqrt{	
\frac{ \Rey^{*}}{\Rey_{0}}	
}.
\label{Re_ratio}
\eeq
from a solution at Reynolds number $\Rey$. The scaling would be exact
if the walls were infinitely far away. In the presence of the walls, the 
scaled states can be taken as initial conditions in a Newton refinement
and ECS on the new scales can be obtained.

For the numerical simulations we use 
Gibson's \textit{Channelflow}-code \cite{Gibson2009b} and
the optimized Newton methods for the determination of ECS.
As a starting point, we use two equilibrium solution that are similar to Eq1 and Eq7 of
\cite{Gibson:2009kp}, which differ in their vortical content. They will collectively be
referred to as EQ when they are turned into equilibrium solutions in the center
and as TW when they are scaled as travelling waves near walls.

\section{Families of scaling solutions in plane Couette flow}

\begin{figure}
\begin{center}
\includegraphics[]{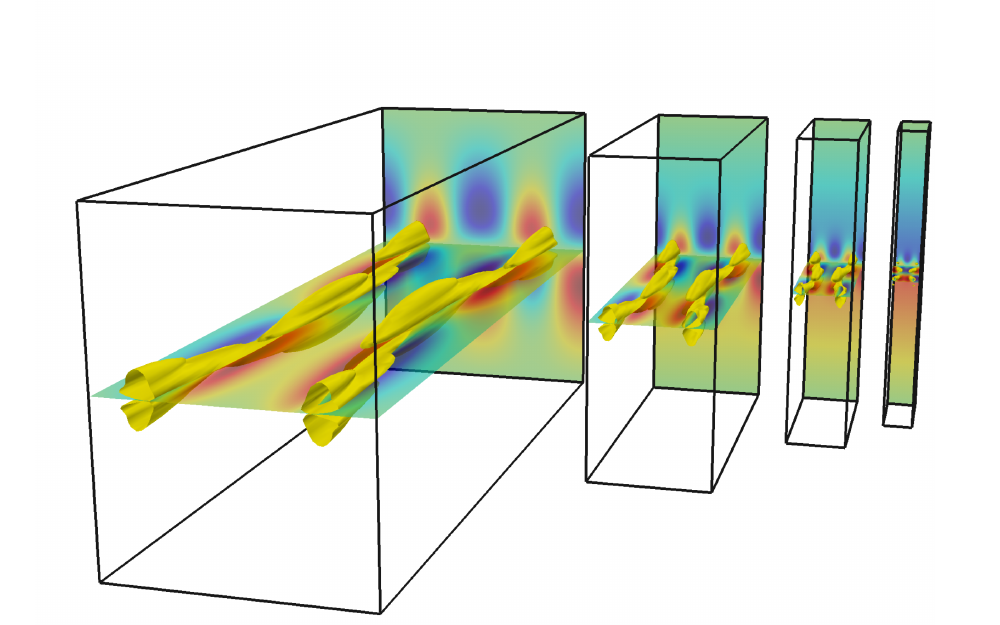}
\end{center}
\caption{Visualization of the exact solutions localized in the center of the channel for 
Reynolds number $100,000$.  The four solutions are scaled in width and length
by factors of $2$, $4$, and $8$. The yellow surfaces show iso-contours of the Q-vortex criterion. The streamwise velocity component is
color coded from low values (blue) to high values (red) in the mid plane
cutting through the structure and in the plane at the back of the domain.
\label{fig:u_center}}
\end{figure}

\subsection{Stationary states in the center}
We begin with solutions that are localized in the center of the domain and stationary
so that $c=0$. The initial computational domain has spanwise and streamwise periodicity of 
$0.5 \pi$ and $1.5\pi$, respectively, and a height of $2$.
Consistent with previous analysis, we can determine the state accurately with a resolution 
$N_{x} \times N_{y} \times N_{z}=32 \times 97 \times 48$ near $Re^*=650$. 
For higher Reynolds numbers, the resolution has to be increased, 
e.g. at $Re_0=10^{5}$ we use a resolution 
$N_{x} \times N_{y} \times N_{z}=24 \times 185 \times 56$.
At each $\Rey$ we carefully checked convergence and that the used resolution is sufficient.

Although the rescaling works for any value of $\lambda$, 
we will here study powers of two only.
Thus, using scaling factors $\lambda$ of $2$, $4$ and $8$, we
can identify equilibrium solutions that are reduced in size by factors
$2$, $4$, and $8$ in all three directions. The corresponding 
Reynolds numbers increase by factors of 4, 16, and 64.
The stationary state is initially
identified at Reynolds number near $3000$, and then scaled up in
Reynolds number to higher values up to $192,000$. In order to be able to compare
the states at a prescribed Reynolds number, the states are then traced
to a Reynolds number $\Rey=100,000$. Visualizations for the states
in the center are shown in Figure \ref{fig:u_center}.

\begin{figure}
\begin{center}
\includegraphics[]{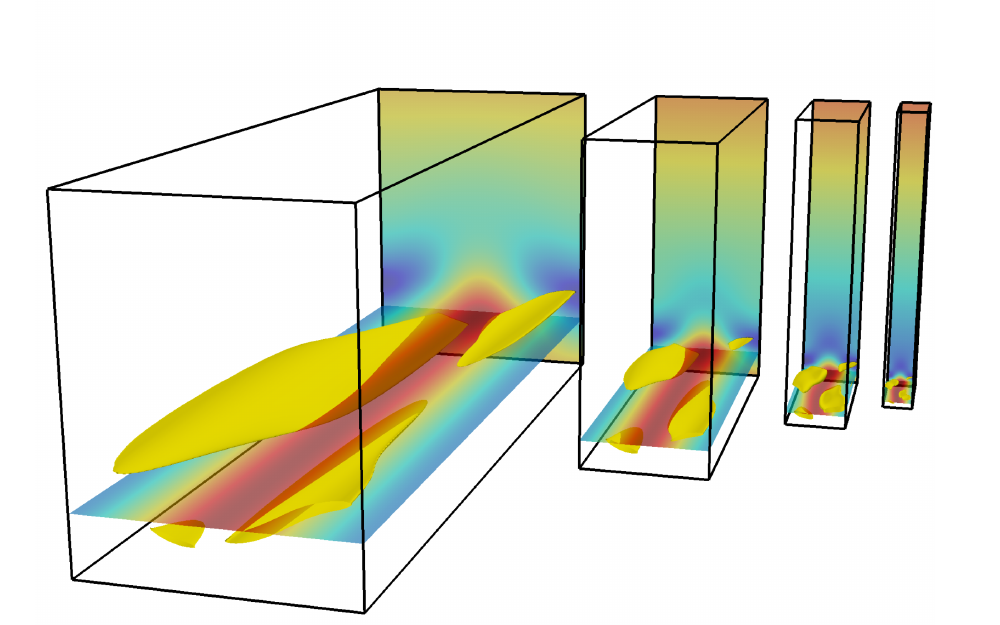}
\end{center}
\caption{Visualization of the lower branch states of exact solutions near the 
lower wall, at Reynolds number $\Rey=100,000$. The representation is the same
as in Fig.÷\ref{fig:u_center}.
\label{fig:u_walls}}
\end{figure}

\subsection{Stationary states near a wall}
For the structures localized in the center, the midplane where $c=0$ is a good point
of reference, and scaling by $\lambda$ moves the boundary planes further away. 
For states close to a wall, the point of reference has to be the wall. 
Eventually, the state will move closer to the wall and the
phase speed will approach $\pm U_0$, the speed of the wall. Accordingly, we
shift the domain upwards by $d$ and change to a co-moving frame of reference
where $U(y=0)=0$ and $U(y=2d)=2U_0$. The equation for the stationary
state remains similar to (\ref{eq:fixedpoint}).

The spanwise and streamwise wavelengths of the initial domain 
are $0.4\pi$ and $0.877\pi$, respectively. The scaled states are shown
in Figure \ref{fig:u_walls}. This family of ECS is reminiscent of the 
structures used in attached  eddy models for the logarithmic layer in wall turbulence,
where the flow field is modeled by a hierarchy of eddies which are attached to the 
wall and whose dimensions increase with the distance to the wall \cite{Woodcock:2015ut}.

\begin{figure}
\begin{center}
\includegraphics[]{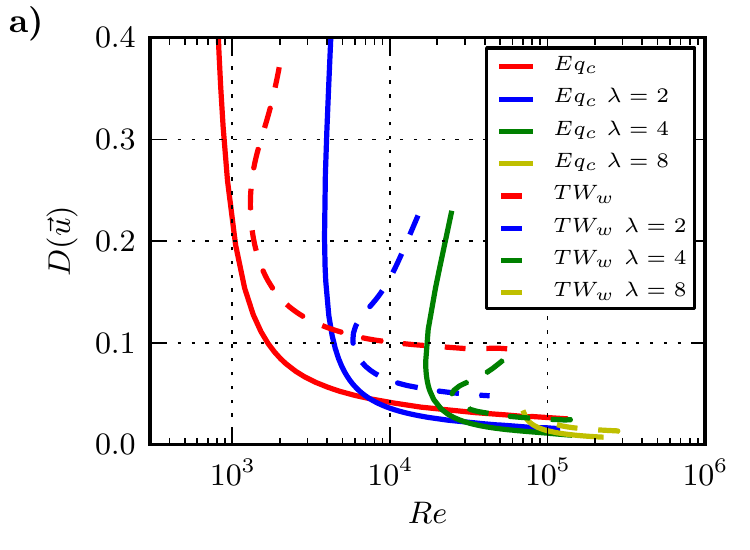}
\includegraphics[]{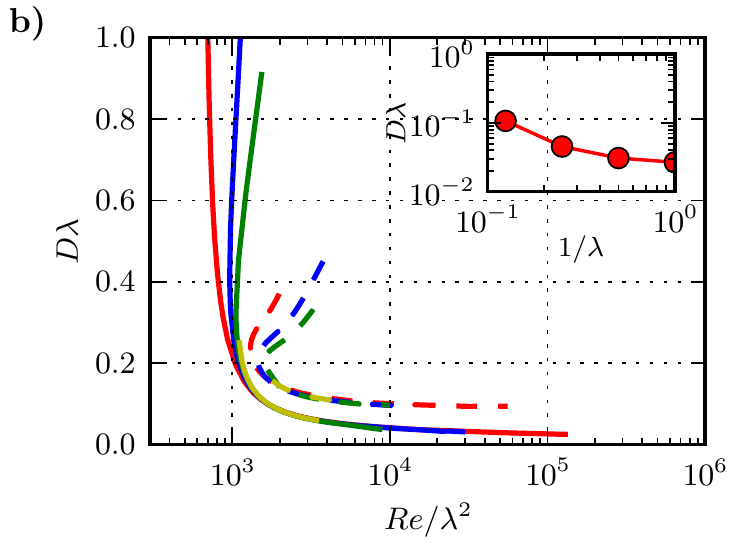}
\end{center}
\caption{Bifurcation diagrams for the scaled ECS in the center
(derived from $EQ$, full lines) and near the wall (derived from $TW$, dashed lines).
(a) shows the unscaled bifurcation diagrams and (b) the rescaled ones where a collapse
of the data is observed. 
The inset in (b) shows the rescaled dissipation of the wall states at $Re=10^5$ 
versus the scaling parameter $\lambda$.
\label{fig_BifDiag1}}
\end{figure}

\subsection{Bifurcation diagrams}
A bifurcation diagram using the volume averaged dissipation,
\beq
D(\mathbf{u})=\frac{1}{
2L_{x}L_{z}} \int_{0}^{L_{z}} \int_{-1}^{1} \int_{0}^{L_{x}} \|\nabla  \times \mathbf{u}\|^{2} 
dx dy dz , \label{Eqn_Diss}
\eeq 
along the ordinate is given in figure \ref{fig_BifDiag1}a.
If one uses the rescaled dissipation on the abscissa and the rescaled Reynolds number 
$\Rey^{*}=\lambda^{2}Re_0$  on the ordinate the bifurcation curves should collapse.
Indeed, in the rescaled bifurcation diagram shown in figure \ref{fig_BifDiag1}b) 
the collapse of the bifurcation curves for different $\lambda$ is very good.

At a fixed Reynolds number the volume averaged dissipation 
(eqn. \ref{Eqn_Diss}) decreases with $\lambda$. 
But the states also become smaller with increasing $\lambda$, filling only
a fraction $1 / \lambda$ of the domain in the wall-normal direction, so that
the rescaled dissipation $\lambda D$ is a better measure for the dissipation
and its scaling. 
At fixed value of $\Rey$ the rescaled dissipation increases with $\lambda$, as shown 
for $TW$ and $Re=10^{5}$  and $1.2\cdot10^{5}$ in the inset in 
Figure \ref{fig_BifDiag1}b).

\begin{figure}
\begin{center}
\includegraphics[]{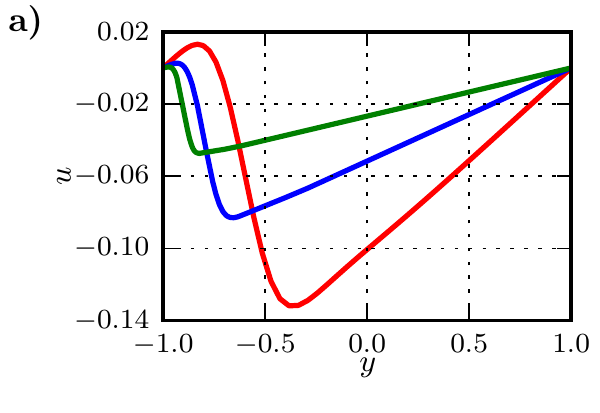}
\includegraphics[]{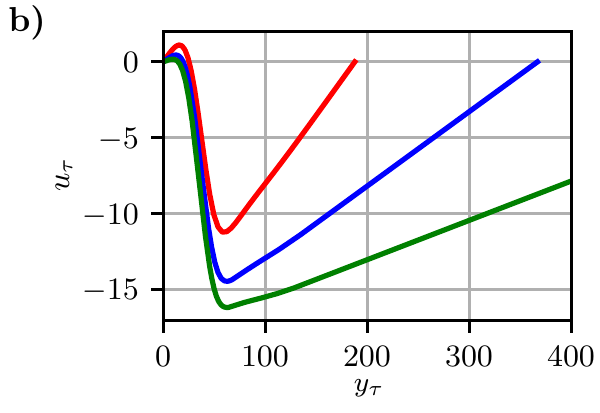}\\
\includegraphics[]{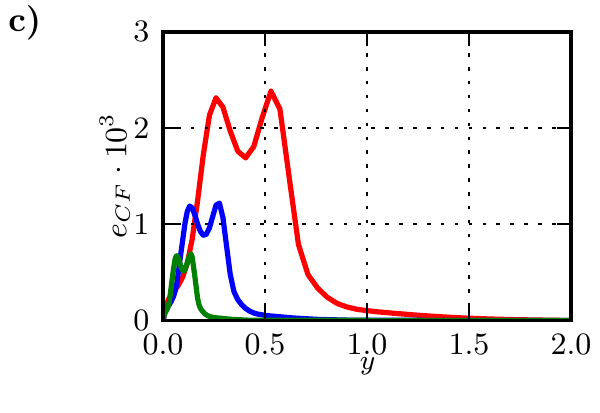}
\includegraphics[]{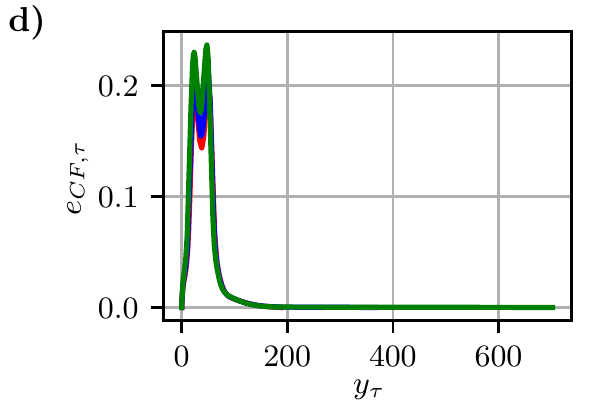}
\end{center}
\caption{Profiles for $TW_{w}$ in the original coordinates (left column) and in 
wall units (right column). The linear laminar profile has been subtracted.
(a) Profile of the streamwise velocity.  (b) Rescaled velocity profile. 
c) Profiles in the cross flow energy density $e_{CF}$ as measure for the location of vortices. 
d) Rescaled profile of $e_{CF}$ 
in wall units.
\label{fig_YLoc}}
\end{figure}

\subsection{Localization properties in the normal direction}
Scaling of the solutions requires that in the normal direction they are not or only weakly 
influenced by the walls. In wall bounded flows, distances and velocities are usually measured
in wall units, based on the viscosity and the friction velocity $u_\tau$. The wall friction is given by
\beq
\tau=\nu \left\langle \frac{\partial u_x}{\partial_y}\right\rangle_{w}
\eeq
where the index $w$ indicates an average at the wall. 
Then the units for velocity are
$ u_{\tau}= \sqrt{\tau} $ and $ \ell_\tau=\nu/u_\tau $. With the scaling of the ECS given by
(\ref{u_scaled}) and the scaling of the viscosity, one finds that the scales at two 
Reynolds numbers $\Rey^{*}$ and $\Rey_0$ (as in (\ref{Re_ratio})) are related by
\beq
\tau^{*} = \frac{1}{\lambda^2} \tau_0 \qquad 
u_\tau^{*} = \frac{1}{\lambda}u_\tau^{(0)} \qquad
\ell_\tau^{*} = \frac{1}{\lambda}  \ell_\tau^{(0)} 
\eeq
Therefore, the rescaling of velocities and lengths by $\lambda$ is equivalent to a rescaling
to wall units if the solutions scale exactly. Since we have to adjust the solutions a little bit in order
to obtained converged states at the respective Reynolds numbers, there are small deviations
in the friction factors, and hence in the wall units. Specifically, for the cases shown here,
$l_\tau$ varies between $3.02\cdot 10^{-3}$ for the largest state with $\lambda=1$
and $3.13\cdot 10^{-3}$ for the smallest state with $\lambda=8$, all evaluated at
Reynolds number $\Rey=100,000$. 

The mean downstream velocity the profiles for the wall states are shown in the top row
of Figure \ref{fig_YLoc}. With increasing $\lambda$ the states become ever more 
localized near the wall and the maximal amplitude becomes smaller. However,
from the maximum to the upper wall, the decay is very slow and essentially linear,
as shown in the rescaled solution in the right column. 

Other measures provide a much clearer signal for the localization.
For instance, the cross flow energy density,
\begin{equation}
e_{CF}(y)=\frac{1}{L_{x}L_{z}}\int_{0}^{L_{x}}  \int_{0}^{L_{x}} 
\left(v^{2} + w^{2} \right) dx dz ,
\end{equation}
which is shown in Figure \ref{fig_YLoc}c, decays much faster outside the region where
the vortices are located. The rescaled curves for the cross flow energy density 
(Figure \ref{fig_YLoc}d) collapse perfectly  and reveal the similarity of the solutions. 
In the normal direction, the ECS modify the mean velocity within the structure,
but do not provide any forces further away. In the absence of forces, the laminar shear
profile is linear, which shows that the linear profile in the outer region is a consequence
of the viscous mediation between the downstream velocity at the outer edge of the ECS
and the velocity at the upper wall.

In the other directions, one can adapt the model for streamwise localization in plane 
Couette flow \cite{Brand:2014he,Barnett:2016ty} to show
that ECS are exponentially localized in the streamwise direction. In the spanwise
direction, the decay seems to be  somewhat stronger, as also noticed for large
scale ECS \cite{Schneider:2010id}. 

\begin{figure}
\begin{center}
\includegraphics[]{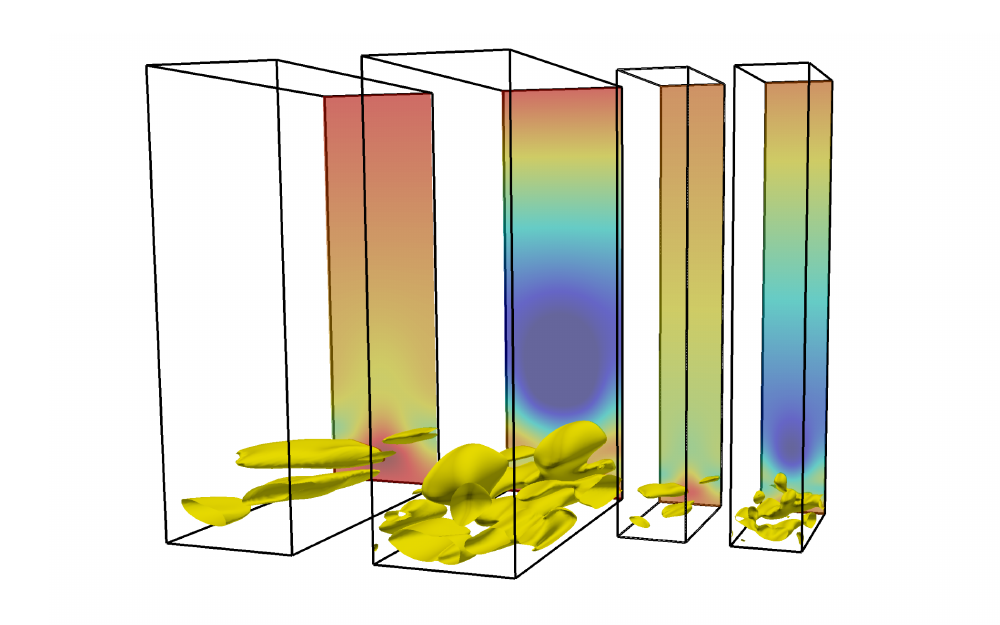}
\end{center}
\caption{Comparison of the lower (a,c) and upper (b,d) branch states for the solutions 
localized near the lower wall. For the left pair of figures (a,b) the Reynolds number 
is $18,200$ and for the right pair (c,d) it is $52,000$.
The yellow surfaces are iso-contours of the Q-vortex criterion.  
The used value of Q is $0.002$ for (a,b) and $Q=0.005$ for (c,d).  
In the back-plane the streamwise velocity component is color-coded. 
The minimal (blue) and maximal (red) values of $u$ are $-0.3$ and $0.05$ in (a,b) and 
$-0.16$ and $0.05$ in (c,d), respectively.}
\label{fig:upper_lower_comparison}
\end{figure}

The upper branch states for both solutions have a much larger wall-normal 
extension than the lower branches states 
(see Figure \ref{fig:upper_lower_comparison}).
%
%
Thus, they are more strongly influenced by the wall which causes an
imperfect scaling, especially for low values of $\lambda$.
For larger $\lambda$ the range of the upper branch states in the wall normal 
direction decreases, resulting in better scaling.

For both the states in the center and near the wall, the lower branch shows 
less variation in streamwise direction with increasing distance to the bifurcation point, 
which is a common feature of lower branch states \cite{Wang2007,Gibson2014}.

\subsection{Stability properties}
\begin{figure}
\begin{center}
\includegraphics[]{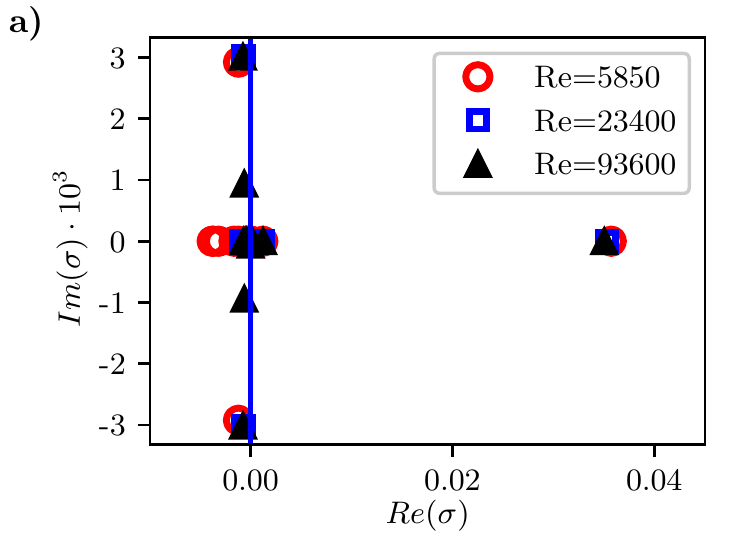}
\includegraphics[]{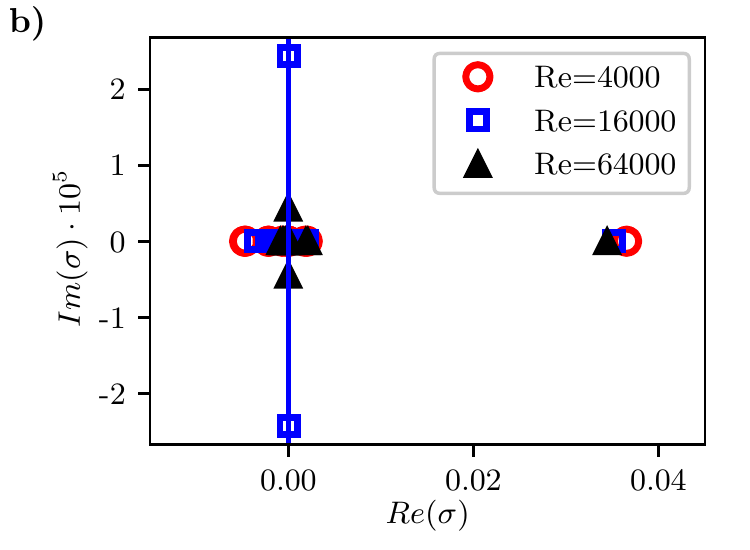}
\end{center}
\caption{Stability properties of the rescaled ECS.
(a) Real and imaginary parts of the leading eigenvalues for 
$EQ$ at $Re=5850$ and the scaled solutions for 
$\lambda=2$ and $4$ at $Re=23400$ and $Re=93600$.
b) Leading eigenvalues for $TW$ at $Re=4000$ and the scaled solutions 
for $\lambda=2$ and $4$ at $Re=16,000$ and $Re=64,000$.\label{fig_EV}
}
\end{figure}

In order to analyze the stability properties of the scales states
the eigenvalues of the ECS are calculated in computational 
domains with spatial periodicities equal to those of the states.
The leading eigenvalues are shown in Figure \ref{fig_EV}.
All states are unstable, but the number of unstable eigenvalues is rather low.
The result show 
that the leading eigenvalues of states whose Reynolds numbers differ by the 
scaling factor $\lambda$,  the leading eigenvalues are almost identical.
This is a consequence of the scaling in (\ref{eq:fixedpoint}), which leave
the time-dependence invariant.
Thus, the dynamics close to the ECS is similar in the adjusted domains.

\section{Spanwise and streamwise localization}
In addition to the ECS in domains that are periodic in the downstream and spanwise direction,
we also tracked states that are localized in these directions 
\cite{Schneider:2010id,Schneider:2010jz}.
As in other cases, good initial guesses can be be obtained
by applying suitable window functions to extract nuclei for localized states
from spatially extended states \cite{Gibson2014}.
We demonstrate this for one streamwise and two spanwise localized equilibrium solutions 
which are related to $EQ$ and $TW$. 

\begin{figure}
\begin{center}
\includegraphics[]{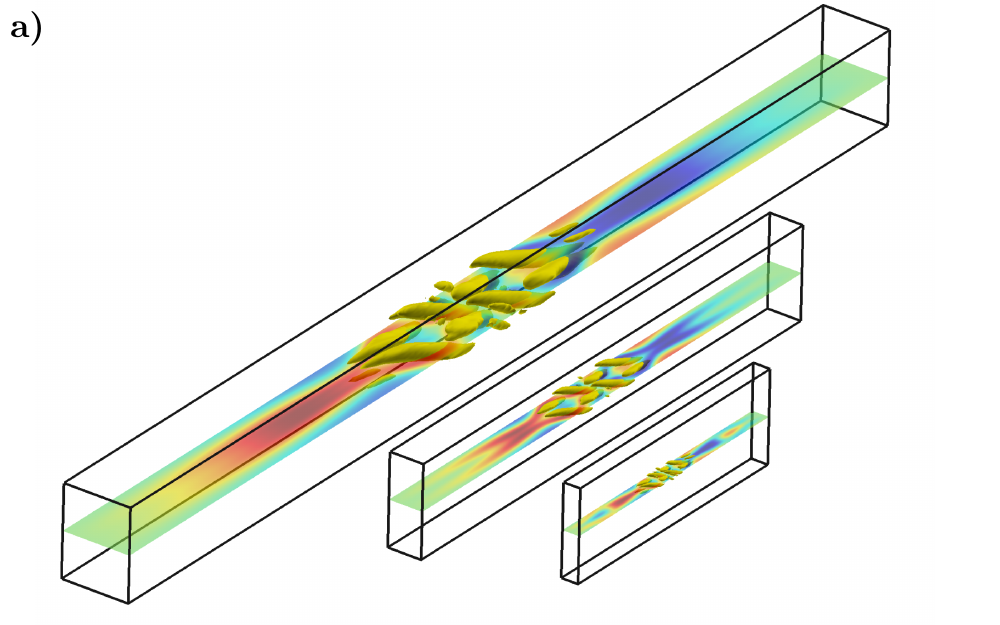}\\
\includegraphics[]{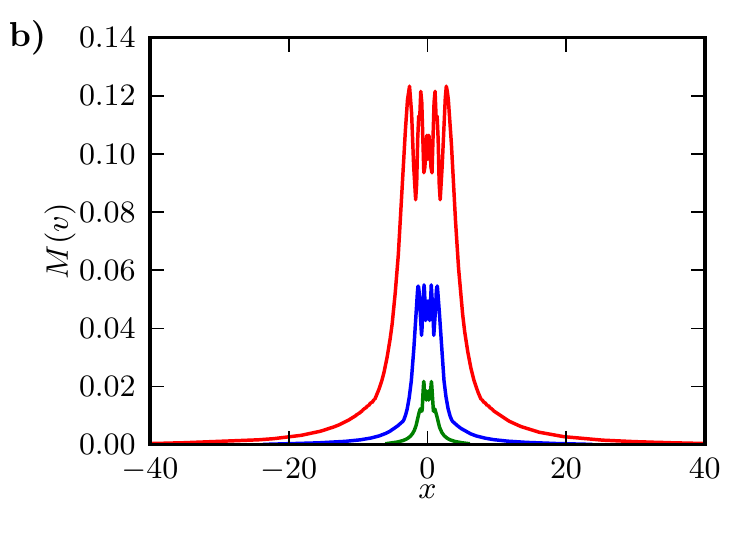}
\includegraphics[]{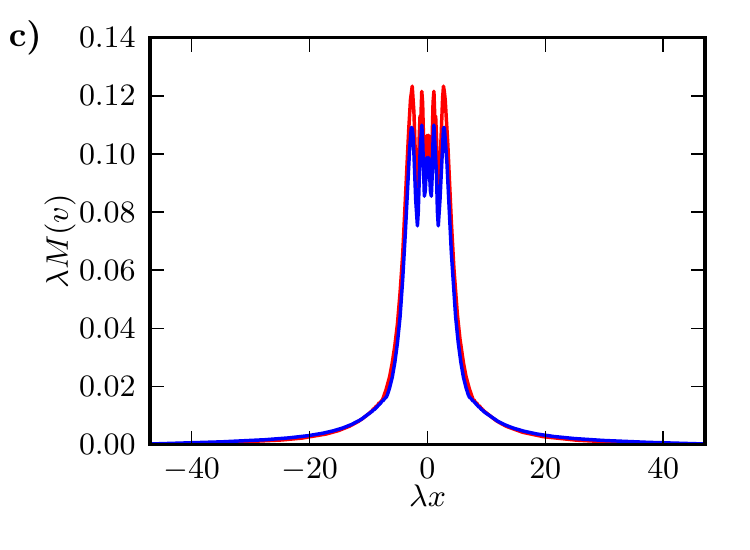}
\end{center}
\caption{Streamwise localized equilibrium solutions.
(a) Visualizations of the flow fields at Reynolds numbers  $900$, $3750$, and  $15600$. 
The streamwise and spanwise lenghts of the largest domain are 
$7.5\pi$ and $0.5\pi$, respectively.
(b) Maximum  of the wall-normal velocity component $M(v)$ versus the 
streamwise position $x$ for the three localized equilibrium solutions.
(c) $M(v)$ rescaled by $\lambda$. Note that the green line is covered by the blue one,
showing the almost perfect scaling for high Reynolds number.
\label{fig_XLocStates}}
\end{figure}

Figure \ref{fig_XLocStates} shows  visualizations of a streamwise localized equilibrium
state related to $EQ$, and of the corresponding 
scaled solutions. As for the streamwise extended solutions, the
visualizations for the different values of $\lambda$ look very similar because the 
scaling works quite well. 
The bifurcation diagram of the streamwise localized states is more complicated 
than for the spatially extended states.
In particular, it has not been possible to trace the family of scaled states to a common 
fixed value of $\Rey$. They are therefore visualized at different Reynolds numbers
that differ approximately by a factor $\lambda^2=4$.

The figure shows that the vortex tubes are oriented in a V-shape which is also a 
feature of the recently identified doubly localized equilibrium states
of PCF \cite{Brand2014}.  
The states show an exponential decay of the velocity components in their tails.
Models for the streamwise decay length $\ell$ in an exponential representation
$u\propto \exp(-|x|/\ell)$ of the downstream variation show that $\ell$
increases with Reynolds number but decreases with spanwise wavelength
\cite{Brand2014,Zammert2016c}. For the rescaled ECS studied here this means
that the stretching of $\ell$ due to the increase in Reynolds number is
compensated by the reduction of the lateral scales so that the overall
all directions can be rescaled by $\lambda$.

\section{Conclusions}
The tracking of ECS from large scales to ever smaller scales at increasing
Reynolds numbers show that a multitude of small-scale ECS  populate
the state space of flows at high Reynolds numbers. Their localization in the
wall-normal direction show that similar states can also appear in shear flows
with curvature in the mean profile, such as Poiseuille flow, 
since eventually the states will only probe the local shear 
gradient \cite{Deguchi:2015gu}.

The two cases studied here are located at the midplane of the domain,
where the mean velocity is $0$, and near the walls, where the mean
advection speed approaches the speed of the wall. For states at some distance
to the wall, one can keep that distance fixed and scale the states so that
they become localized at that height. Initially, for low Reynolds numbers,
there will be some influence of the walls, but then for higher Reynolds numbers 
and more localized states, the influence from the walls will become smaller, and 
one can anticipate that the states become similar to the ones in the center.

The localization in the normal direction, and also in the other
directions, implies that sets of states can be combined to form ECS of more
complex spatial structures: for superpositions of localized states the 
nonlinear terms are weak if they are very far apart, and even if the 
interactions between the two states are stronger, the superposition 
provides a good starting point for a Newton method. 
In the few cases where we 
attempted such superpositions, the Newton method was able to adjust
the flow fields so that converged ECS that are have two (or more) centers
of localization could be obtained. 

For the staggered attached eddies used to represent boundary layers
\cite{Perry:1991ab,Perry:1994ab}
a simple superposition will not work because the states overlap not only in their
fringes but in their core. The interactions will then be more complicated
than the simple perturbative adjustment that worked for spatially separated
structures, and remains a challenge for computations. 
Without that interaction, the wall states can be used to form 
approximate hierarchical superpositions of structures of the type
discussed in \cite{Woodcock:2015ut}. Therefore,
the wall states described here are a promising starting point for modelling
hierarchical structures near walls. 

The similarity of the stability properties of the rescaled ECS in the rescaled domain 
suggests that the frequency with which ECS are visited, which is directly related to 
their instability, is preserved
under scaling. Therefore, the small scale ECS should be visited and be visible
in the flows on that scale as frequently as on the large scale. Indeed, DNS simulations
in narrow domains \cite{Yang:2017up} show structures
that are similar to the ones described here. In an extensive data analysis of
homogeneous shear flows, \cite{Dong:2017cd} deduced structures
consisting of vortex and streaks which they termed roller states. They are
similar to one half of the ECS shown in
Figure \ref{fig:u_center}. Two such states can be combined to form a stationary
states, which is consistent with the observation that the ECS are stationary
states, whereas the roller states of Dong et al are transient. Nevertheless,
the similarity between ECS and observations in DNS is encouraging 
and shows that it is possible to detect ECS not only in low-Re transitional
flows \cite{Hof:2004ab,Schneider:2007ib,Kerswell:2007ds} 
but also in high-Reynolds number situations. It should therefore also
be possible to extend the use of ECS for the characterization of transitional 
flows to fully developed turbulent flows and to provide, for instance,
a dynamical basis for the attached eddy hypothesis.

Finally, we note that the structures described here should also be observed in the
presence of curved walls: when the Reynolds number increases the curvature
becomes small on the scale of the structures and hence negligible. So very close to the
wall in pipe flow, or in G\"ortler flow, similar structures should appear.
 
\section*{Acknowledgements}
We thank the participants of the 2017 KITP Workshop
"Recurrent Flows: The Clockwork Behind Turbulence" for discussions, and
the National Science Foundation for partial support of KITP 
under Grant No. NSF PHY11-25915.
This work was also supported in part by the Deutsche Forschungsgemeinschaft within 
FOR 1182 and by Stichting voor Fundamenteel Onderzoek der Materie (FOM) within 
the program "Towards ultimate turbulence".

\section*{References}


\end{document}